\documentstyle[12pt]{article}
\begin{document}
\tolerance=100000
\baselineskip=20pt

\def\gsim{ \lower .75ex \hbox{$\sim$} \llap{\raise .27ex \hbox{$>$}} }
\def\lsim{ \lower .75ex \hbox{$\sim$} \llap{\raise .27ex \hbox{$<$}} }
\def\mincir{\raise -2.truept\hbox{\rlap{\hbox{$\sim$}}\raise5.truept
\hbox{$<$}\ }}
\def\magcir{\raise -2.truept\hbox{\rlap{\hbox{$\sim$}}\raise5.truept
\hbox{$>$}\ }}

%
\def\ea{{et al. }}
\def\aa{{ A\&A}}
\def\aas{{ A\&AS}}
\def\araa{{ ARA\&A}}
\def\apj{{ ApJ}}
\def\apjl{{ ApJ}}
\def\apjs{{ ApJS}}
\def\aj{{ AJ}}
\def\apl{{Ap. Lett.}}
\def\mnras{{ MNRAS}}
\def\pasp{{ PASP}}
\def\ajpas{{Aust. J. Phys. Ap. Suppl.}}
\def\baas{{ BAAS}}
\def\phrev{{Phys. Rev.}}
\def\rmph{{Rev. Mod. Phys.}}
\def\spscirev{{Space Sci. Rev.}}
\def\nat{{ Nat}}
\def\bul{{ Bull. AAS}}

\begin{center}

 THE NEAR--IR--OPTICAL--UV EMISSION OF \\
 BL LACERTAE OBJECTS$^a$

\vspace{ 1.5cm}


{\sc E. Pian$^1$, R. Falomo$^2$, R. Scarpa$^3$, and A. Treves$^1$ }

\vspace{ 1.2cm}

\end{center}

\begin{description}

\item
$^1$ Scuola Internazionale Superiore di Studi Avanzati, via Beirut 2--4,
34014  Trieste, Italy

\item
$^2$ Osservatorio Astronomico di Padova, v. Osservatorio 5, 35122, Padova,
Italy

\item
$^3$ Dipartimento di Astronomia dell'Universit\`a di Padova,
v. Osservatorio 5, 35122 Padova, Italy

\end{description}

\vspace{1.2cm}
\begin{center}
To appear in The Astrophysical Journal
\end{center}

\vfill
$^a$ Based on observations obtained at the European Southern Observatory,
La Silla, Chile, and with the International Ultraviolet Explorer collected at
the ESA Tracking Station at Villafranca

\newpage

\baselineskip=0.9truecm

\begin{center}
ABSTRACT
\end{center}
\vspace{0.5cm}

 Near--infrared, optical and ultraviolet quasi--simultaneous
 observations
 of 11 BL Lacertae objects are reported. For all but one
 source the dereddened
 spectral flux distribution in the $8\cdot10^{13}-2\cdot10^{15}$ Hz
 frequency range can be described by a single power law
 f$_\nu \propto \nu^{-\alpha}$ with average spectral index
 $<\alpha>$ = 0.88 $\pm$ 0.42 (standard deviation) plus, where relevant,
 the contribution of the host galaxy. In most cases the non simultaneous soft
 X--ray fluxes obtained by the {\it Einstein Observatory} lie on or below the
 extrapolation of the power law.
 The results are compared with the average
 spectral properties of other samples of BL Lacs studied separately in
 the IR--optical and in the UV bands. The implications for existing models
 of the objects are shortly discussed.

\vspace{0.5cm}

\noindent
 Subject headings: BL Lacertae objects: general --- infrared: galaxies
--- ultraviolet: galaxies --- multifrequency observations: galaxies

\newpage

\begin{center}
{\small 1. INTRODUCTION }
\end{center}

\vspace{0.5cm}

The spectral flux distribution (SFD) of BL Lacertae objects (BL Lacs)
  from the far--IR to
X--ray range is supposedly dominated by non thermal emission due
to synchrotron radiation with possible contribution, at the higher
frequencies, of the Compton mechanism.
The existence of breaks or of spectral curvature is
important for constructing models for acceleration and radiation of
electrons, and for constraining the emitting region, in particular to
ascertain if it has an anisotropic structure where relativistic beaming
possibly takes place.

The overall SFD is generally described either by broken power laws
or by continuous steepening (see {\it e.g.} Cruz--Gonzalez \& Huchra 1984;
Landau \ea 1986; Ghisellini \ea 1986; Brown \ea 1989; Ballard \ea 1990).
Most of previous papers refer to non simultaneous observations, not
systematically corrected for extinction due to dust inside our
Galaxy and for the thermal contribution produced by the host galaxy starlight.
 The stellar contribution, if
non negligible with respect to the non thermal
emission, produces a steepening of the energy
distribution in the optical and a flattening in the
near--IR, while reddening extinction
introduces a steepening of the
continuum mainly at optical--UV frequencies. For instance, the
spectral break observed in some objects between near--IR
and optical is completely removed when proper
reddening corrections are applied (see {\it e.g.} Falomo \ea 1993b).

We report here on quasi--simultaneous ($\Delta$t $\mincir$ 1 day) near--IR,
optical and UV  ($8\cdot10^{13}-2\cdot10^{15}$ Hz) observations of
11 BL Lacs obtained in the course of our 10 years systematic multifrequency
study of blazars (see Falomo \ea 1993a,b,c and references therein).
Most of them have been previously published, but are here reconsidered
and analysed using a homogeneous procedure which takes into account
the galactic extinction and the host galaxy contribution.
The main aim of the paper is to derive the intrinsic shape of the
non thermal emission from near--IR to UV frequencies, based on a relatively
extended sample, though not complete from the statistical point of view.

The plan of the paper is as follows:  In \S 2 the modes of observation
and of data analysis are described. In \S 3 the general characteristics
of the near--IR--optical--UV SFDs are outlined and their
extrapolations are compared with non simultaneous far--IR (IRAS) and soft
X--ray ({\it Einstein}) data.
In \S 4 the results are discussed and the
spectral slopes of the 11 considered objects are compared with those of two
samples of 33 and 24 objects observed, respectively, in the near--IR--optical
 and in the UV ranges.

\vspace{1cm}

\begin{center}
{\small 2. OBSERVATIONS AND DATA ANALYSIS}
\end{center}

\vspace{0.5cm}

The target objects are reported in Table 1,
sorted by increasing right ascension (Col. 1), along with dates of
observations (Col. 2), redshift (Col. 3), galactic extinction values
(Col. 4, see below). Col. 5 labels whether the source is radio--strong (RS)
or radio--weak (RW), following the criterion
of Ledden \& O'Dell (1985). RS objects substantially
correspond to radio selected ones, and RW to X--ray selected ones.
A journal of observations is given in Table 2, where fluxes in selected
bands are reported.
All observations were obtained with the same instrumentation and analysed
using a uniform procedure.
For most objects, observations in the near--IR, optical and UV bands were
taken within $\sim$1 day. In two cases (2005--489 and 2155--304)
we combined simultaneous IR--optical and optical--UV observations
taken at different epochs to study the IR--optical--UV emission (see
next section).
For two sources (0048--097 and 0422+004), more than one
overall spectrum has been obtained, allowing a comparison of the
spectral shape at different states.

\bigskip
\begin{center}
{\small 2.1 \it {Near--IR Photometry}}
\end{center}
\bigskip

J,H,K and L photometry was obtained at the European Southern Observatory
(ESO) 2.2m telescope (+ InSb photometer).
A 15 arcsec circular aperture with chopper throw of 20 arcsec in
the E--W direction was used. Statistical 1--$\sigma$ errors are less than 0.1
mag in all bands. Conversion to flux units is made according to the
zero--magnitude fluxes given in Bersanelli, Bouchet \& Falomo (1991).

\bigskip
\begin{center}
{\small 2.2 \it {Optical Spectrophotometry }}
\end{center}
\bigskip

Optical spectrophotometry of the sources was gathered at the ESO
1.5m telescope equipped with a Boller and Chivens
spectrograph and CCD detector.
Spectra were taken at a resolution of $\approx$
15 {\rm \AA} \ (~FWHM~) through a long slit of 8 arcsec width. Standard
reduction procedures were applied to obtain flux calibrated spectra. From
repeated observations of standard stars (Stone 1977; Baldwin \& Stone 1984)
during each night, we derived a photometric accuracy better than 10\%. To
increase the signal--to--noise ratio, we averaged the fluxes binning
the spectra over bands of 100 {\rm \AA}.

\bigskip
\begin{center}
{\small 2.3 \it {Ultraviolet Spectra }}
\end{center}
\bigskip

UV spectra derive from the Short Wavelength Primary (SWP;
range: 1200--1950 {\rm \AA}) and the Long Wavelength Primary (LWP; range:
2000--3200 {\rm \AA}) cameras onboard of the International Ultraviolet
Explorer (IUE). IUE line--by--line images were flux calibrated  using
curves provided by Bohlin \& Holm (1980) and Bohlin \ea (1990) for the SWP
camera and Cassatella, Lloyd \& Gonzalez--Riestra (1989) for the LWP camera.
Net spectra were
extracted using an implementation of the Gaussian extraction procedure
(GEX, Urry \& Reichert 1988) running within the MIDAS interactive
analysis system (see Falomo \ea 1993a).
We added a 10\%  photometric error to the statistical flux errors.

The UV spectra of 2005--489 and 2155--304 were retrieved from the ULDA
archive.
We note that IUE spectra extracted with the IUESIPS procedure, show
systematic flux differences by 5 to 10 \% with respect to extraction with
the SWET method of Kinney, Bohlin \& Neill (1991).
In order to avoid such flux distortion due to different extraction methods,
we have applied correction on IUESIPS extraction based on comparison with
the UV data reported by Edelson \ea (1992).

Overall data were corrected for galactic interstellar reddening using the
extinction values listed in Table 1. They were deduced from the
hydrogen column density (Stark \ea 1992) assuming
N$_H$/E$_{B-V}$ = $5.2 \cdot 10^{21}$ (Shull \& Van Steenberg 1985) and
R = 3.1 (Rieke \& Lebofsky 1985).
The interstellar extinction curve of Cardelli, Clayton \& Mathis (1989)
was used.

\vspace{1cm}

\begin{center}
{\small 3. SPECTRAL FLUX DISTRIBUTIONS }
\end{center}
\vspace{0.5cm}

A composite SFD was constructed for each
object from quasi--simultaneous IR, optical and UV observations.
For many objects, we have some more optical observations, spaced of
$\sim$ 1 day from the
one presented here for a given epoch (see Falomo, Scarpa \&
Bersanelli 1994). Because no significant flux variations during day time
scales are found, we do not expect that intra--day variability substantially
 affects our results.

For all sources we attempted a decomposition of the SFD in
terms of a power law plus the contribution of a host galaxy (see Fig. 1).
As a model for the starlight
contribution we assumed the energy distribution of a standard giant elliptical
galaxy (Yee \& Oke 1978) for the optical region, extended to the near--IR using
the colors for ellipticals given by Arimoto \& Yoshii (1987). To subtract the
host galaxy component, the redshift of the object is needed. This is not
available for 2 objects, therefore we assumed a rough
estimate of $z$ based on fit optimization.
Parameters of the decomposition are: the spectral index $\alpha$ of the power
law component and the percentage $P_{g}$ of starlight contribution at
5500 {\rm \AA}.
We performed a least squares fit of the SFDs and report in Table 1 the
best fit values. Col. 9 gives the estimated absolute magnitude of the
host galaxy.

For most sources, after proper correction for the galactic
extinction and subtraction of the starlight component, the IR--optical--UV
non thermal emission is adequately well fitted by a single power law.
In four objects (0048--097, 0118--272, 1538+149, 1553+113) the host galaxy is
not detectable, thus no decomposition of the SFD was performed.
The SFDs of the first three objects are well described by a single power law
(see Fig. 1 and Table 1). For 1553+113 a single power law does not
account for the SFD but a considerable discontinuity is apparent
($\Delta\alpha \sim 0.8$) at $\sim$ 3000 {\rm \AA}, which makes this source
the only case where a marked spectral change is observed (see also
Falomo \& Treves 1990). For seven objects we observe a bending of the SFD,
which we ascribe to the host galaxy. After decomposition of the thermal
component, the data are satisfactorily represented by a single power law.
In two cases (0323+022 and the high state of 0422+004) the fit is however
poor ($\chi^2 \sim 2$).

For two sources (0048--097 and 0422+004) we obtained SFDs at two different
epochs. As for 0048--097, the two SFDs differ by a factor 1.5 in the optical
range and a slightly steeper spectrum ($\Delta\alpha \sim 0.3$) is exhibited
in the lower state (see Fig. 1).
In the low state the UV data appear systematically
below the power law fit, suggesting a possible further steepening of the
SFD in the UV.
However, due to the low flux level of this IUE spectrum, its steepening
must be regarded with caution.

Similarly, 0422+004 shows an optical flux variability of a
factor $\sim 1.5$. After subtraction of the host galaxy, the
non thermal component is described at both epochs by a single power law.
No significant changes of spectral index are found here (see Table 1).
The source had been observed also in a previous occasion
(1987, Falomo \ea 1989), but the lack of a correction for reddening and for
the host galaxy contribution led to the interpretation of the 1987 and
1988 SFDs in terms of broken power laws or of logarithmic parabolas.

For 2005--489 and 2155--304 we do not have IR, optical and UV simultaneous
observations, therefore we combined
simultaneous
IR--optical observations (Falomo \ea 1993b) with simultaneous optical--UV
observations obtained at different epochs.
Optical--UV observations were scaled to match near--IR--optical data
in the optical range. This is justified by the property that the optical
 spectral
index remains rather constant, even if the flux level changes sizeably
with time (Falomo et al. 1994).
The combined SFD of 2005--489 (see Table 2 and Fig. 1) exhibits a clear
bending in the near--IR due to host galaxy. The decomposition of the SFD
in terms of power law plus host galaxy is not very satisfactory
 ($\chi^2 \sim 3$),
because far--UV data appear to be systematically below the fit.
A similar situation occurs in  the case of 2155--304, although the bending
is here less evident.
Also in this case, far--UV fluxes lie systematically below the overall
fit. This, similarly to the case of 1553+113, might be due to an intrinsic
spectral break of
the power law component at log($\nu$) $\sim$ 15.1.
Although an intrinsic spectral steepening in the UV band of these two
sources appears an acceptable
explanation, we believe that other causes as IUE flux extraction and
calibration, proper reddening correction and the lack of a single epoch
observation may build up a similar effect.
Based only on these data, we cannot draw a firm conclusion on this point.

In order to extend our SFDs to adjacent bands (far--IR and X--rays) we
collected from the literature IRAS data (Impey \& Neugebauer 1988) and
soft X--ray fluxes gathered by the {\it Einstein Observatory}
(Owen, Helfand \& Spangler 1981; Madejski \& Schwartz 1983;
Ulmer et al. 1983; Ledden \& O'Dell 1985; Worrall \& Wilkes 1990;
Elvis et al. 1992).
The IRAS fluxes represent in four cases only upper limit values.
The overall (non simultaneous) SFDs from far--IR to X--rays
of our target objects are reported in Fig. 2.
We note that the extrapolation of our IR--to--UV power law fit to
far--IR is consistent with IRAS fluxes (and upper limits)
taking into account the variability regime of the objects.
On the other hand, in all cases but 0521--365 the observed X--ray flux
lies on or below our extrapolations.

\vspace{1cm}
\begin{center}
{\small 4. DISCUSSION }
\end{center}
\vspace{0.5cm}

The main result of the paper is that in all of the examined objects but
1553+113 the SFD in the IR--optical--UV band is properly represented by a
unique
power law, when corrections for reddening and host galaxy are taken into
account. This is at variance with previous results, which found a turnover
in blazar spectra between IR and UV wavelengths ({\it e.g.} Cruz--Gonzalez \&
Huchra 1984; Ghisellini \ea 1986) or a smooth curvature from the radio
to the UV domain ({\it e.g.} Landau \ea 1986; Brown \ea 1989). We believe that
most of the difference is due to the use of simultaneous observations and
proper corrections for extinction and for the stellar contribution. This is
illustrated by the case of 0422+004, for which the proper account of the
severe galactic extinction and of the host galaxy contribution, determines
a radical change in the description of the emission components.
Due to the higher, and more reliable, value of the extinction coefficient
adopted here for 0323+022, the fit
decomposition parameters for this object are slightly different
than those previously obtained (Falomo et al. 1993a).

Our decomposition of the SFD allows to estimate also the
starlight contribution of program objects.
The derived absolute magnitudes correspond to an $8 \times 8$ arcsec$^2$
effective aperture.  They are in general good agreement with
those obtained from direct imaging studies, considering the aperture
correction, which is crucial for low redshift objects
(see {\it e.g.} the cases of
 0323+022 studied by Filippenko et al. 1986; 0521--365, Keel 1986;
0414+009, Falomo \& Tanzi 1991; 2155--304, Falomo, Pesce \& Treves 1993).
Furthermore, recent optical imaging of 0301--243 confirmed the
presence of a diffuse nebulosity surrounding the object (Falomo 1994, in
preparation).

The average spectral index of the power law component in our sample
is $<\alpha>$ = 0.88 $\pm$ 0.42 (standard deviation), with
a marked tendency for RW objects to be flatter than RS
ones at $\sim 3 \sigma$ probability level; the average spectral slopes are
in fact, respectively, $<\alpha>$ = 0.45 $\pm$ 0.15 (s.d.) and
$<\alpha>$ = 1.17 $\pm$ 0.22 (s.d.).
Comparison with the results from two broader samples of BL Lacs supports
our finding that the average
spectral slope  in the IR--optical--UV band has no discontinuities.
We refer to a list of 33 objects (Falomo \ea 1993b), whose simultaneous
IR--optical data were properly
corrected for reddening and host galaxy contribution.
The second sample has been considered by Pian \& Treves (1993) containing
 24 BL Lacs simultaneously observed in the two IUE bands.
In Table 3 we report the average spectral indices of the different
samples, which appear very similar. It
is noticeable that also in the broader samples RW objects have
average harder spectra than RS ones.
This subdivision, which is known since some time, was tested with the
Kolmogorov--Smirnov method at the 99\% significance level by Pian \& Treves
(1993) and received further observational
support (Giommi, Ansari \& Micol 1993).
In the framework of jet emission models, it can be interpreted as related to
a different beaming angle in the two classes, and to a decreasing opening
angle of the jet with decreasing emission frequency and increasing distance
from the nucleus. RS BL Lacs are more beamed toward us, and then
present a more relativistically enhanced lower frequency emission than
RW ones (Maraschi \ea 1986; Celotti \ea 1993;
Maraschi, Ghisellini \& Celotti 1993).

The absence of breaks or curvature in the spectral shape in
the $8\cdot10^{13}$ to
$2\cdot10^{15}$ Hz frequency interval suggests that a single mechanism is
responsible for the radiation in the whole range. The obvious
candidate is thin synchrotron radiation, as specifically indicated by
the optical and UV polarization.
 We note that the observed X--ray fluxes lie
systematically on or below the extrapolation of the non thermal emission.
Although  the X--ray data were not taken simultaneously, this is indicative
that no component in addition to the synchrotron one is present in the
soft X--rays, and in particular the synchrotron self--Compton process is
probably negligible. Instead, the data suggest a steepening of the spectrum
toward soft X--ray frequencies. This means
that the electron acceleration mechanism
at the higher energies has a typical time scale longer than the electron
lifetimes, so that injection of high energy particles is
less efficient than radiative losses.

Finally, we note that both IR--optical and UV observations of BL Lacs
indicate that the spectral slope changes only modestly with the flux level
(Falomo \ea 1993b; Falomo \ea 1994; Edelson 1992; Urry \ea 1993).
This fact, together with our results, suggests that the whole
IR--optical--UV slope is unique and constant even when the intensity
varies by large factors. Such lack of spectral variability does not arise
in a natural way in any of the proposed models for BL Lacs, apart from those
implying gravitational lensing (Ostriker \& Vietri 1985; 1990).
This very constancy seems to us a key constraint in a theoretical progress of
our understanding of BL Lac emission.

\newpage

We have reconsidered in this work the results of a 10 years observing program
realized in collaboration with P. Bouchet, L. Chiappetti, L. Maraschi and
E.G. Tanzi, who are here gratefully acknowledged. We thank the anonymous
referee for many useful suggestions.
This research has made use of the graphics package Super Mongo by R. Lupton
and P. Monger.

\newpage
\hoffset=0cm
\begin{center}
{\small REFERENCES}
\end{center}
\begin{description}


\item
Arimoto, N., \& Yoshii, Y. 1987, \aa, 173, 23

\item
Baldwin, J. A., \& Stone, R. P. S. 1984, \mnras, 206, 241

\item
Ballard, K. R., Mead, A. R. G., Brand, P. W. J. L., \& Hough, J. H. 1990,
\mnras, 243, 640

\item
Bersanelli, M., Bouchet, P., \& Falomo, R. 1991, \aa, 252, 854


\item
Bohlin, R. C., \& Holm, A. V. 1980, IUE NASA Newsletter, 10, 37

\item
Bohlin, R., Harris, A. W., Holm, A. V., \& Gry, C. 1990, ApJS, 73, 413

\item
Brown, L. M. J., \ea 1989, \apj, 340, 129

\item
Cardelli, J. A., Clayton, G. C., \& Mathis, J. S. 1989, ApJ, 345, 245

\item
Cassatella, A., Lloyd, C., \& Gonzalez--Riestra, R. 1989,
IUE ESA Newsletter, 35, 225

\item
Celotti, A., Maraschi, L., Ghisellini, G., Caccianiga, A., \& Maccacaro, T.
1993, \apj, 416, 118

\item
Cruz--Gonzalez, I., \& Huchra, J. P. 1984, \aj, 89, 441

\item
Edelson, R., Pike, G. F., Saken, J. M., Kinney, A., \& Shull, J. M. 1992,
\apjs, 83, 1

\item
Edelson, R. 1992,
\apj, 401, 516

\item
Elvis, M., Plummer, D., Schachter, J., \& Fabbiano, G. 1992, ApJS, 80, 257

\item
Falomo, R., Bouchet, P., Maraschi, L., Tanzi, E. G., \& Treves, A. 1988,
\apj, 335, 122

\item
Falomo, R., Bouchet, P., Maraschi, L., Tanzi, E. G., \& Treves, A. 1989,
\apj, 345, 148

\item
Falomo, R., \& Treves, A. 1990, \pasp, 102, 1120

\item
Falomo, R., \& Tanzi, E. G. 1991, \aj, 102, 1294


\item
Falomo, R., Treves, A., Chiappetti, L., Maraschi, L., Pian, E., \&
Tanzi, E. G. 1993a, \apj, 402, 532

\item
Falomo, R., Pesce, J. E., \& Treves, A. 1993, ApJ, 411, L63

\item
Falomo, R., Bersanelli, M., Bouchet, P., \& Tanzi, E. G. 1993b,
\aj, 106, 11

\item
Falomo, R., Pian, E., Scarpa, R., \& Treves, A. 1993c, in
Proc. of IAU Symp. 159, Active Galactic Nuclei across the Electromagnetic
Spectrum, in press

\item
Falomo, R., Scarpa, R., \& Bersanelli, M. 1994, \apjs, in press

%

\item
Filippenko, A. V., Djorgovski, S., Spinrad, H., \& Sargent, W. L. W. 1986,
\aj, 91, 49


\item
Ghisellini, G., Maraschi, L., Tanzi, E.G., \& Treves, A. 1986,
\apj, 310, 317

\item
Giommi, P., Ansari, S.G., \& Micol, A. 1993, in
Proc. of IAU Symp. 159, Active Galactic Nuclei across the Electromagnetic
Spectrum, in press



\item
Impey, C. D., \& Neugebauer, G. 1988, \aj, 95, 307

\item
Keel, W. C., 1986, ApJ, 302, 296

\item
Kinney, A. L., Bohlin, R. C., \& Neill, J. D. 1991, PASP, 694, 103

\item
Landau, R., \ea 1986, \apj, 308, 78

\item
Ledden, J. E., \& O'Dell, S. L. 1985, ApJ, 298, 630

\item
Madejski, G. M., \& Schwartz, D. A. 1983, ApJ, 275, 467

\item
Maraschi, L., Ghisellini, G., Tanzi, E. G., \& Treves, A. 1986, ApJ, 310, 325

\item
Maraschi, L., Ghisellini, G., \& Celotti, A. 1993, in
Proc. of IAU Symp. 159, Active Galactic Nuclei across the Electromagnetic
Spectrum, in press

\item
Ostriker, J. P., \& Vietri, M. 1985, Nature, 318, 446

\item
Ostriker, J. P., \& Vietri, M. 1990, Nature, 344, 45

\item
Owen, F. N., Helfand, D. J., \& Spangler, S. R. 1981, ApJ, 250, L55

\item
Pian, E., \& Treves, A. 1993, \apj, 416, 130


\item
Rieke, G. H., \& Lebofsky, M. 1985, \apj, 288, 618

\item
Shull, J. M., \& Van Steenberg, M. E. 1985, \apj, 294, 599


\item
Stark, A. A., Gammie, C. F., Wilson, R. W., Bally, J., Linke, R. A.,
Heiles, C., \& Hurwitz, M. 1992, ApJS, 79, 77

\item
Stone, R. P. S. 1977, \apj, 218, 767

%
%
%

\item
Ulmer, M. P., Brown, R. L., Schwartz, D. A., Patterson, J., \&
Cruddace, R. G. 1983, \apj, 270, L1


\item
Urry, C. M., \& Reichert, G. 1988, IUE NASA Newsletter, 34, 96

\item
Urry, C. M., \ea 1993, \apj, 411, 614

\item
Worrall, D. M., \& Wilkes, B. J. 1990, ApJ, 360, 396

\item
Yee, H. K. C., \& Oke, J. B. 1978, \apj, 226, 753

\end{description}

\newpage

\baselineskip=20pt
%
%
\hoffset=-1.5cm
\begin{center}
\begin{tabular}{llccccrcc}
\multicolumn{9}{c}{\small TABLE 1} \\
\multicolumn{9}{c}{\sc Blazars Observed in Near--IR--Optical--UV}\\
& & & & & & & & \\
\hline
\hline
& & & & & & & & \\
\multicolumn{1}{c}{Object} & \multicolumn{1}{c}{Date} & $z^a$ & $A_V$ &
RS/RW & $\alpha^b$ & \multicolumn{1}{c}{$P_g^c$} &
$\chi^2$ & M$_V^d$ \\
\multicolumn{1}{c}{{\small (1)}} & \multicolumn{1}{c}{{\small (2)}} &
{\small (3)} & {\small (4)} & {\small (5)} & {\small (6)} &
\multicolumn{1}{c}{{\small (7)}} & {\small (8)} & {\small (9)} \\
& & & & & & & & \\
\hline
& & & & & & & & \\
$0048-097$& 87 Jan 8  &$\cdots$& 0.22 & RS & $0.93 \pm 0.06$ & $\cdots$ &
0.87 & $\cdots$\\
          & 88 Aug 3  &        &      &   & $1.20 \pm 0.06$ & $\cdots$ &
1.39 & $\cdots$\\
$0118-272$& 89 Aug 10 & 0.559  & 0.09 & RS & $1.20 \pm 0.03$ & $\cdots$ &
0.40 & $\cdots$\\
$0301-243$& 89 Aug 9 & \multicolumn{1}{r}{(0.2)} & 0.11 & RS & $0.79 \pm 0.06$
 & 11 & 0.90 & --22.2\\
$0323+022$& 89 Aug 10 &0.147   & 0.50 & RW & $0.23 \pm 0.12$ & 30 &
1.92 & --22.2\\
$0414+009$& 89 Feb 15 &0.287   & 0.51 & RW & $0.54 \pm 0.15$ & 3 & 0.06 &
 --21.3 \\
$0422+004$& 88 Jan 10 & \multicolumn{1}{r}{(0.1)} & 0.42 & RS & $1.20 \pm 0.15$
& 10 & 1.88 & --21.5\\
          & 89 Feb 13 &        &      &   & $1.19 \pm 0.15$ & 15 &
0.70 & --21.5\\
$0521-365$& 87 Jan 8  &0.055   & 0.21 & RS & $1.43 \pm 0.09$ & 51 &
1.39 & --21.7\\
$1538+149$& 88 Aug 2  &0.605   & 0.20 & RS & $1.34 \pm 0.06$ & $\cdots$ &
0.23 & $\cdots$\\
$1553+113^e$& 88 Aug 2  &$\cdots$& 0.22 & RW &  & $\cdots$ &  & $\cdots$\\
$2005-489$& 86 Sep/89 Aug&0.071   & 0.33 & RW & $0.57 \pm 0.06$ & 12 &
3.21 & --22.2\\
$2155-304$& 84 Nov/89 Aug & 0.116   & 0.10 & RW & $0.47 \pm 0.06$ & 3 & 0.94
& --22.2 \\
& & & & & & & & \\
\hline
& & & & & & & & \\
\multicolumn{9}{l}{$^a$ Values in parentheses are from
the best fit optimization procedure.} \\
\multicolumn{9}{l}{$^b$ Errors correspond to 3--$\sigma$ uncertainties.} \\
\multicolumn{9}{l}{$^c$ Percentage of host galaxy contribution at 5500
{\rm \AA} (observer frame). } \\
\multicolumn{9}{l}{$^d$ Values are not K--corrected and
computed assuming H$_0$=50 km s$^{-1}$ Mpc$^{-1}$; q$_0$=0.} \\
\multicolumn{9}{l}{$^e$ IR--Optical region:
$\alpha=0.78\pm0.06$ ($\chi^2=0.38$); UV region: $\alpha=1.56\pm0.12$
($\chi^2=0.21$). } \\
\end{tabular}
\end{center}
%

\hoffset=-1.5cm
\begin{center}
\begin{tabular}{cccccccccc}
\multicolumn{10}{c}{\small  TABLE 2} \\
\multicolumn{10}{c}{ \sc Journal of Observations }\\
& & & & & & & & & \\
\hline
\hline
& & & & & & & & & \\
 Name & MJD$^a$ & F$_K^b$ & $\Delta$t$^c$ & F$_V^d$ & F$_{LWP}^e$ &
 $\Delta$t$^c$ & F$_{SWP}^f$ & $\Delta$t$^c$ & Refs.$^g$ \\
& & & & & & & & & \\
 \hline
& & & & & & & & & \\
 $0048-097$
 & 46804.074 & 10.38 & --0.016 & 2.31 & 1.23 & 0.986   & 0.59 & --0.046 & 1 \\
 & 47377.336 & 6.98  &   0.012 & 1.51 &      &         & 0.25 & 1.092 & 2,3 \\
 & & & & & & & & & \\
 $0118-272$
 & 47748.289 & 6.71  & 0.078   & 1.25 & 0.50 & --0.120 &      &       & 4 \\
 & & & & & & & & & \\
 $0301-243$
 & 47749.277 & 4.32  & 0.066   & 1.16 & 0.53 & --1.985 & 0.31 & --1.062 & 4 \\
 & & & & & & & & & \\
 $0323+022$
 & 47749.367 & 2.65  & --0.012 & 0.67 & 0.26 & --1.011 & 0.15 & 0.906 & 4 \\
 & & & & & & & & & \\
 $0414+009$
 & 47570.000 & 2.31  & 4.100   & 0.64 & 0.26 & 4.917   & 0.19 & 3.856 & 5 \\
 & & & & & & & & & \\
 $0422+004$
 & 47171.078 & 16.41 & --0.020 & 2.38 &      &         & 0.24 & 0.898 & 6 \\
 & 47571.031 & 12.79 & 0.012   & 1.66 & 0.33 & 1.899   & 0.19 & 0.827 & 2,3 \\
 & & & & & & & & & \\
 $0521-365$
 & 46803.152 & 18.20 & 0.043   & 2.38 &      &         & 0.17 & 1.754 & 2,3 \\
 & & & & & & & & & \\
 $1538+149$
 & 47377.035 & 2.93  & --0.031 & 0.39 & 0.12 & 0.384   &      &       & 4 \\
 & & & & & & & & & \\
 $1553+113$
 & 47376.969 & 28.31 &   0.023 & 8.43 & 3.44 & 0.348   & 1.37 & --0.552 & 7 \\
 & & & & & & & & & \\
 $2005-489$
 & 46683.102 &       &         & 8.93 & 4.11 & 0.291   & 1.97 & 0.171 &
2,3,8 \\
 & 47748.105 & 26.79 & 0.000   & 8.32 &      &         &      &       &
2,3,8 \\
 & & & & & & & & & \\
 $2155-304$
2,3,8 \\
2,3,8 \\
 & 46017.051 &       &         & 24.49 & 1.18 & --0.047 & 0.98 & 0.107 &
2,3,8 \\
 & 47749.156 & 26.55 & --0.008 & 12.33 &     &         &      &       &
2,3,8 \\
& & & & & & & & & \\
\hline
& & & & & & & & & \\
\multicolumn{10}{l}{$^a$ Modified Julian Day of the optical observation.} \\
\multicolumn{10}{l}{$^b$ Flux at 2.19 $\mu$m in mJy. } \\
\multicolumn{10}{l}{$^c$ Time lag with respect to the optical observation
(days).} \\
\multicolumn{10}{l}{$^d$ Flux at $\sim$ 5500 {\rm \AA} in mJy. } \\
\multicolumn{10}{l}{$^e$ Flux at $\sim$ 2800 {\rm \AA} in mJy. } \\
\multicolumn{10}{l}{$^f$ Flux at $\sim$ 1400 {\rm \AA} in mJy. } \\
\multicolumn{10}{l}{$^g$ References to papers where some of the data
have been presented:} \\
\multicolumn{10}{l}{ $~~$ (1) Falomo \ea 1988. (2) Falomo \ea 1993b.
(3) This paper.} \\
\multicolumn{10}{l}{ $~~$ (4) Falomo \ea 1993a. (5) Falomo \& Tanzi 1991.
(6) Falomo \ea 1989. } \\
\multicolumn{10}{l}{ $~~$ (7) Falomo \& Treves 1990. (8) Falomo, Scarpa \&
Bersanelli 1994.} \\

\end{tabular}
\end{center}


\newpage
\hoffset=-0.8cm
\begin{center}
\begin{tabular}{lccc}
\multicolumn{4}{c}{\small TABLE 3} \\
\multicolumn{4}{c}{\sc  Average Spectral Indices$^a$}\\
&&&\\
\hline
\hline \\
& $<\alpha_{IROP}>$ & $<\alpha_{UV}>$ & $<\alpha_{IROPUV}>$ \\
&  Falomo \ea 1993b  & Pian \& Treves 1993 & This paper \\
&&&\\
\hline \\
All             & 1.08 $\pm$ 0.34 (33) & 0.97 $\pm$ 0.41 (24) & 0.88 $\pm$
0.42 (10) \\
Radio--weak & 0.68 $\pm$ 0.28  (8) & 0.66 $\pm$ 0.30 (10) & 0.45 $\pm$
0.15 (4) \\
Radio--strong  & 1.20 $\pm$ 0.30 (25) & 1.20 $\pm$ 0.33 (14) & 1.17 $\pm$
0.22 (6) \\
&&&\\
\hline
&&&\\
\multicolumn{4}{l}{$^a$ The quoted errors represent the standard
deviations from the mean quantities} \\
\multicolumn{4}{l}{$~~$ and values in parentheses refer to
the number of objects. } \\

\end{tabular}
\end{center}


\newpage

\begin{center}
{\small FIGURE CAPTIONS}
\end{center}

\vspace{1.6cm}

{\sc Figure 1}

\vspace{0.3cm}

\noindent
Spectral flux distributions of BL Lac objects observed simultaneously at
near--IR, optical and UV frequencies. Data (filled squares) are
corrected for galactic interstellar extinction. The solid line is the best
fit model, which is either a single power law or the combination of a host
galaxy (dotted line) and a power law component (dashed line).
Open squares represent the spectrum of the standard elliptical after
rebinning in conformity to the observed data points.

\vspace{0.5cm}

{\sc Figure 2}

\vspace{0.3cm}

\noindent

Overall spectra from far--IR to soft X--rays. Comparison of the
non thermal component (dashed line) with non simultaneous far--IR and
X--ray data (open squares).
Near--IR--optical--UV data from our dereddened observations are also
reported (filled squares).

\end{document}